\documentclass[nofootinbib,aip,graphicx]{revtex4-1}
\usepackage[final]{graphicx}
\usepackage{epsfig}
\usepackage{amssymb}
\usepackage{amsmath}
\usepackage{epstopdf}
\usepackage{subfloat}
\usepackage{lineno}
\usepackage{float}
\usepackage{color}
\usepackage{units}
\usepackage{enumerate}
\usepackage{multirow}
\usepackage{rotating}
\usepackage{array}
\usepackage[version=4]{mhchem}
\usepackage{chemist}
\usepackage{tikz,chemfig}
\usepackage{blkarray}
\usepackage{gensymb}
\usepackage{amsmath,amssymb,mathrsfs,bm,amsthm}
\usepackage{array}
\usepackage{mathtools}
\usepackage{float}
\usepackage[caption = false]{subfig}
\usepackage{ulem}

\newcommand\arcbetweennodes[3]{%
    \pgfmathanglebetweenpoints{\pgfpointanchor{#1}{center}}{\pgfpointanchor{#2}{center}}%
    \let#3\pgfmathresult}

\newcommand\arclabel[6][red,-stealth,shorten <=1pt,shorten >=1pt]{%
    \chemmove{%
        \arcbetweennodes{#4}{#3}\anglestart
        \arcbetweennodes{#4}{#5}\angleend
        \ifdim\anglestart pt>\angleend pt \pgfmathsetmacro\anglestart{\anglestart-360}\fi
        \draw[#1]([shift=(\anglestart:#2)]#4)arc[start angle=\anglestart,end angle=\angleend,radius=#2];%
        \pgfmathsetmacro\anglestart{(\anglestart+\angleend)/2}%
        \node[shift=(\anglestart:#2+1pt)#4,anchor=\anglestart+180,inner sep=0pt,outer sep=0pt]at(#4){#6};%
    }%
}

\usepackage{mathrsfs,amsmath,amssymb,mathtools} 

\begin{document}

\title{Intramolecular Vibrational Redistribution in Formic Acid and its Deuterated Forms}
\author{Antoine Aerts} 
\affiliation{Universit\'e libre de Bruxelles, Spectroscopy, Quantum Chemistry and Atmospheric Remote Sensing (SQUARES), 50, av. F. Roosevelt CP 160/09, 1050 Brussels, Belgium}
\email{antoine.aerts@ulb.be}
\affiliation{Department of Chemistry, University of Alberta, Edmonton, Alberta, Canada, T6G 2G2}
\author{Alex Brown}
\affiliation{Department of Chemistry, University of Alberta, Edmonton, Alberta, Canada, T6G 2G2}
\author{Fabien Gatti}
\affiliation{ISMO, Institut des Sciences Mol\'eculaires d'Orsay-UMR 8214 CNRS/Universit\'e Paris-Saclay, F-91405 Orsay, France}

\begin{abstract}
\underline{\textbf{Abstract}}\linebreak
The intramolecular vibrational relaxation dynamics of formic acid and its deuterated isotopologues is simulated on the full-dimensional potential energy surface of Richter and Carbonni\`ere [F. Richter and P. Carbonni\`ere, J. Chem. Phys. \textbf{148}, 064303 (2018)] using the Heidelberg MCTDH package. Mode couplings with the torsion coordinate capturing most of the \textit{trans}-\textit{cis} isomerisation are identified in the dynamics of artificially excited vibrational mode overtones. The \ce{C-O} stretch bright vibrational mode is coupled to the out-of-the plane torsion mode in \ce{HCOOH}, where this coupling could be exploited for laser-induced \textit{trans}-to-\textit{cis} isomerisation. Strong isotopic effects are observed: deuteration of the hydroxyl group, \textit{i.e.}, in \ce{HCOOD} and \ce{DCOOD}, destroys the \ce{C-O} stretch to torsion mode coupling whereas in \ce{DCOOH}, little to no effect is observed.

\end{abstract}

\maketitle

\section{Introduction}
Formic acid (HCOOH) is the simplest molecule that contains a carboxylic acid functional group and it is also one of the simplest molecules that exhibits isomerism between two planar structures. Besides the chemistry-oriented interests in its reactivity, formic acid serves as a benchmark molecule for isolated systems of this class both in theoretical studies\cite{jcp_126_164305,jpca_120_9815,jcp_6_064303,jcp_152_024305,jms_385_111617} and experiments.\cite{cs_formic,pccp_22_25492,jcp_154_064301} Formic acid is found in the Earth's atmosphere\cite{rg_37_227,acp_10_10047,acp_15_6283,grl_47_e2019GL086239} and in the interstellar medium, where it was first detected in 1971 by Zuckerman \textit{et al.}\cite{aj_163_L41} Its spectral and dynamical characteristics are of particular interest given the existence of two conformational isomers, \textit{i.e.}, \textit{trans}- and \textit{cis}-formic acid. The \textit{cis} form is higher in energy than the \textit{trans} isomer by about 1413 cm\textsuperscript{-1} when comparing their zero-point-energies.\cite{jcp_6_064303}
The \textit{trans} isomer is therefore about 800 times more abundant than the \textit{cis} isomer at room temperature, which makes its detection in the gas phase challenging. Using a long absorption path, and the fact there is a much larger dipole moment of the \textit{cis} isomer (as compared to the \textit{trans}), enabled the detection of \textit{cis}-formic acid with microwave spectroscopy by Hocking in 1976.\cite{natur_31_1113} Observations of \textit{cis}-formic acid remained very scarce\cite{jms_216_259,jms_795_49} until recent major improvements using Raman Jet spectroscopy by Suhm, Meyer and coworkers.\cite{jcp_149_104307,cs_formic,pccp_22_25492,jcp_154_064301}

The conformational state of the carboxyl (\ce{-COOH}) group can have significant effects in macromolecules and can influence the conformation of folded proteins.\cite{nc_9_1,nat_337_476,ncb_3_619,nat_329_6136,nat_329_266} This functional group form strong inter- and intramolecular hydrogen bonds that stabilise molecular structures and influence reactivity.\cite{cr_83_83} Such interactions also enabled the recent observation of both conformers of formic acid in aqueous solution at room temperature by Giubertoni \textit{et al.}\cite{jpcl_10_3217}

The isomerisation mechanism of formic acid from the \textit{trans} to the \textit{cis} isomer has already drawn attention in the past.\cite{jms_385_111617,jacs_119_11715,jacs_125_4058,jcp_117_9095,jms_219_191,jpca_46_13346} Intramolecular vibrational redistribution (IVR) should play an important role in the isomerisation mechanism in the gas phase to bring the energy deposited in a bright (infrared-active) vibrational mode, \textit{e.g.}, by laser excitation, towards the torsional degree of freedom that mainly characterises the isomerization pathway. Observations of such an isomerism were limited to matrix-isolated formic acid and most probably assisted by the local environment since the IVR depends strongly on mode couplings. Important tunneling effects were also reported both for the \textit{trans} to \textit{cis} and \textit{cis} to \textit{trans} isomerisations in rare gas matrices. However, in the gas phase, Hurtmans \textit{et al.}\cite{jcp_113_1535} reported local OH bond stretching up to the third overtone (4 $\nu_{\text{OH}}$) in \textit{trans}-formic acid, without evidence of isomerisation. Their subsequent analysis, supported by \textit{ab initio} computations, lead them to suggest a transfer of the hydrogen rather than isomerism following the excitation of formic acid to high energy OH stretching overtones. More recently, Nejad and Sibert\cite{jcp_154_064301} pointed out the mixing between the $\ce{C-O}$ stretch and the torsion involved in the isomerisation mechanism of \textit{trans}-HCOOH; a mixing that is also observed in \textit{trans}-DCOOH.

Almost all fundamentals of formic acid exhibit anharmonic resonances and/or Coriolis interactions.\cite{jcp_126_164305} The latter are beyond the scope of this work, and we present here simulations of IVR by local vibrational excitations with emphasis on the isomerisation process in formic acid. For this purpose, we will use the multiconfiguration time dependent Hartree (MDTDH) method.\cite{mctdh:package,mey90:73,man92:3199,bec00:1,mey03:251,mey09:book} More precisely, we will use the Heidelberg package,\cite{mctdh:package} which requires the Hamiltonian to be in sum-of-product form for efficiency. This requirement was addressed before, where anharmonic resonances are treated explicitly with the full dimensional (9D) potential energy surface (PES) of Richter and Carbonni\`ere.\cite{jcp_6_064303} The PES covers both \textit{trans} and \textit{cis} minima and is valid up to 6000 cm\textsuperscript{-1} above the \textit{trans}-HCOOH zero point energy (ZPE). 
We use a description based on valence polyspherical coordinates.\cite{iun99:3377,gat09:1} 
The kinetic energy operator was generated in these coordinates using the TANA program.\cite{jcp_136_034107,jcp_139_204107,lau02:8560,tana-tnum:package} 
Their validity, both of the potential energy surface and the kinetic energy operator, has been confirmed by a number of studies comparing with experimental vibrational band centers of \textit{trans} and \textit{cis} isomers including deuterated forms.\cite{jcp_6_064303,jcp_152_024305,pccp_22_25492,jcp_154_064301,cs_formic} However, the PES does not permit the study of \ce{OH} (or \ce{OD}) stretching overtones mentioned above ($\geq$ 4 $\nu_{\text{OH}}$), as it has not been fit for these high energy regions.

MCTDH has already been used successfully to describe IVR in tetra-atomic molecules involving motions of large amplitude including isotopic effects, \textit{e.g.}, for HONO \cite{ric04:1306,ric04:6072,ric07:164315} and HFCO.\cite{pas06:194304,pas07:024302,pas08:144304} Recently, Mart\'in Santa Dar\'ia \textit{et al.}\cite{jms_385_111617} have predicted mixed \textit{cis-trans} states that have an energy close to that of the isomerization barrier. However, to the best of our knowledge, no large-amplitude stable torsional vibrational states have ever been reported for formic acid. This is probably due to the high density of states for a system of this size, but also couplings of the torsional out-of-the-plane mode with any of the bright vibrational modes. The purpose of the present study is to identify a possible path towards \textit{trans} to \textit{cis} isomerisation of formic acid in its ground electronic state. 
The challenge, besides the density of states high in energy, is the collective effect of weakly non-resonant ``dark" state(s) (infrared(IR)-inactive, as opposed to ``bright") coupled to the bright states that are expected to be significant.\cite{jcp_99_2261,jpc_100_12735} We will quantify these effects and analyse the redistribution of energy from local vibrational excitations.

\section{Vibrational structure of Formic Acid}
\begin{figure}
\includegraphics[width=0.5\columnwidth]{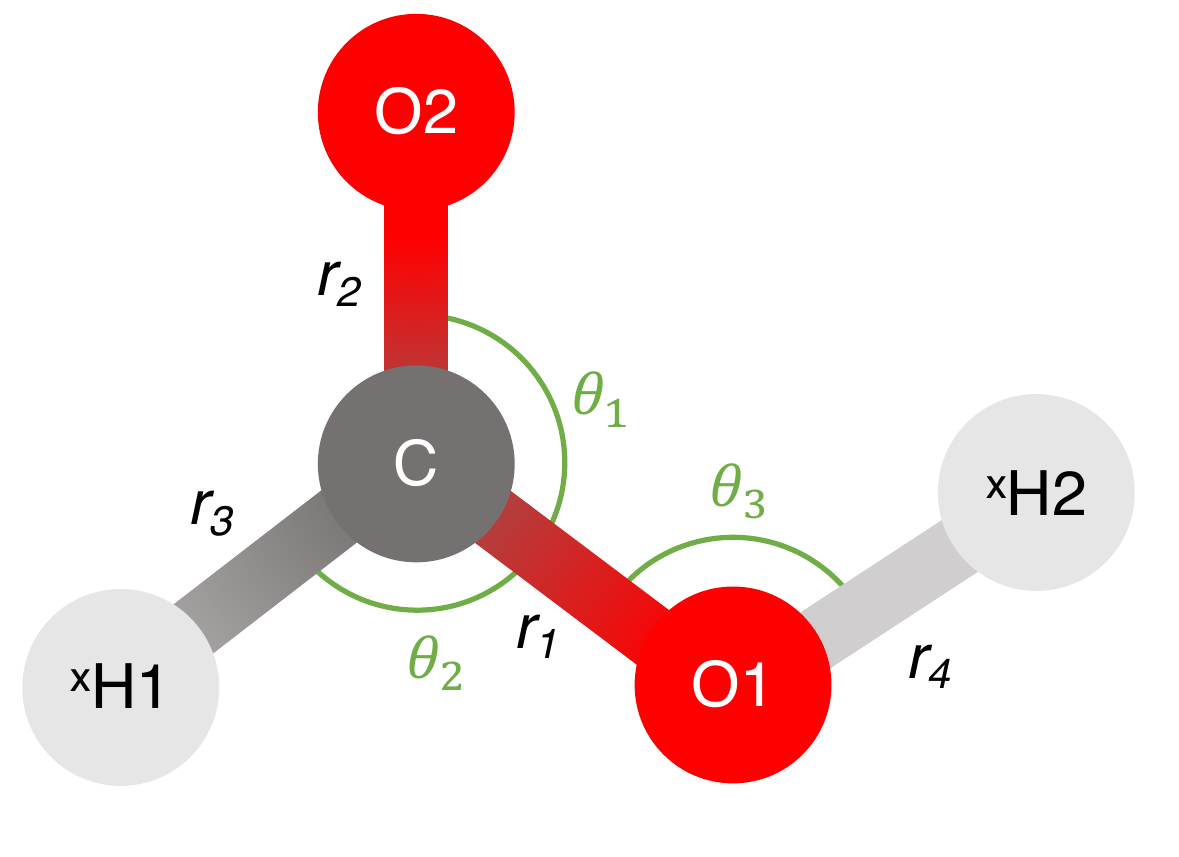}
\caption{Valence coordinates used to describe formic acid; the torsional angles $\tau_1$ and $\tau_2$ are respectively the \textsuperscript{x}H{1} and \textsuperscript{x}H{2} (or D1/D2 depending on deuterium substitution) out of plane coordinates relative to the O{2}CO{1} body.}
\label{fig:zmatrix}
\end{figure}

The vibrational motion of formic acid and its deuterated isotopologues (HCOOH, DCOOD, DCOOH and HCOOD) can be described in terms of 9 modes of vibration. The reported experimental fundamental frequencies are given in Table \ref{table:fundaformic} and compared to computed values using the block-improved relaxation method (see Section~\ref{sec:improvedrelax}) implemented in the Heidelberg MCTDH package on the PES of Richter and Carbonni\`ere.\cite{jcp_6_064303} Note that the ``Herzberg'' vibrational mode notations differ from one species to the other, and therefore, designations based on the explicit atomic motions will be used throughout this work. These descriptions rely on the definition of atoms given in Fig. \ref{fig:zmatrix}.

The low frequency modes allow for several Fermi resonances to occur and, combined with the large density of states, account for difficulties theoretically but also enrich the spectrum of formic acid as recently highlighted by the work of Nejad and Sibert.\cite{jcp_154_064301} 
Isotopic substitution can have a strong effect on the torsional mode ($\tau_2=\tau_{\ce{CO^xH 2}}$) vibrational frequency connected to \textit{trans}-\textit{cis} isomerization and of particular interest here, as shown in Table \ref{table:fundaformic}. The effects of deuteration on the IVR dynamics and therefore on the mode couplings will be analyzed in the present work. 

\begin{table}

\caption{Available experimental (exp.) fundamental vibrational energies (in cm\textsuperscript{-1}) of formic acid and its isotopologues compared to computed values using block-improved relaxation within MCTDH on the PES of Richter and Carbonni\`ere\cite{jcp_6_064303} (MCTDH). Atom labeling is given in Fig.~\ref{fig:zmatrix} Mode assignment is given as $\nu$ = stretching, $\delta$ = bending, $\gamma$ = rocking, $\tau$ = torsion, and def. = deformation.
Error on the last digit(s) is indicated in brackets when published.}
\begin{center}
\resizebox{\textwidth}{!}{%
\begin{tabular}{|c|c|c|c|c|c|c|c|c|c|}
\hline 
\multirow{3}{*}{Assignment}  &  \multicolumn{4}{c|}{HCOOH}   & \multicolumn{4}{c|}{DCOOD}  \\
 & \multicolumn{2}{c|}{\textit{trans}} & \multicolumn{2}{c|}{\textit{cis}} & \multicolumn{2}{c|}{\textit{trans}} & \multicolumn{2}{c|}{\textit{cis}} \\
 &exp.&{\small MCTDH\footnotemark[1]}&exp.&{\small MCTDH\footnotemark[1]}&exp.&{\small MCTDH\footnotemark[2]}&exp.&{\small MCTDH\footnotemark[2]}\\
 \hline 
$\tau_{\ce{CO^xH 2}}$ & 640.73\cite{jms_216_203}&637 & 493.42\cite{jms_795_49}&491 & 493.2254252(35)\cite{jms_193_33} &489 & -- &370 \\
$\tau_{\ce{^xH 1CO}}$ & 1033.47\cite{jms_211_262}&1032 & --&1011 & 873.0\cite{sapa_25_1243} & 871& -- &860 \\
\ce{OCO-CO^xH 2} def. &  626.17\cite{jms_216_203}&623 &--&652  & 554.4394726(50)\cite{jms_193_33}& 552 & -- & 615 \\
$\delta_{\ce{CO^xH 2}}$ & 1306.1\cite{jmst_795_54}&1301 & -- &1246& 945.0\cite{sapa_25_1243} & 944 & 883\cite{pccp_22_25492} & 884 \\
$\gamma_{\ce{C^xH 1}}$ & 1379.05\cite{jms_334_22}&1375 & -- & 1383& 1042\cite{jcp_76_886} & 1035 & --& 1029 \\
$\nu_{\ce{C-O}}$ & 1104.85\cite{jms_211_262} & 1106& 1093\cite{jcp_149_104307,cs_formic} &1097 & 1170.79980(2)\cite{jms_195_324} & 1170 & -- & 1159 \\
$\nu_{\ce{C=O}}$ & 1776.83\cite{jqsrt_110_743} &1774 & 1818\cite{jcp_149_104307,cs_formic} &1810 & 1742.0\cite{sapa_25_1243} & 1759 &  1789\cite{pccp_22_25492}& 1782 \\
$\nu_{\ce{C^xH 1}}$ & 2942.06\cite{cp_283_47} &2937 & 2873\cite{jcp_149_104307,cs_formic}& 2871& 2213.0\cite{sapa_25_1243} & 2228 & 2145\cite{pccp_22_25492} & 2143 \\
$\nu_{\ce{O^xH 2}}$ & 3570.5\cite{cp_283_47}  &3567 & 3637\cite{jcp_149_104307,cs_formic} & 3631& 2631.87379(43)\cite{sapa_56_991} &2629 & 2685\cite{pccp_22_25492} & 2683\\
\hline
\multirow{3}{*}{Assignment} &\multicolumn{4}{c|}{HCOOD}   & \multicolumn{4}{c|}{DCOOH}\\ 
&\multicolumn{2}{c|}{\textit{trans}} & \multicolumn{2}{c|}{\textit{cis}} & \multicolumn{2}{c|}{\textit{trans}} & \multicolumn{2}{c|}{\textit{cis}}\\ 
&exp.&{\small MCTDH\footnotemark[2]}&exp.&{\small MCTDH\footnotemark[2]}&exp.&{\small MCTDH\footnotemark[2]}&exp.&{\small MCTDH\footnotemark[2]}\\
\hline
$\tau_{\ce{CO^xH 2}}$ &508.1320569(45) \cite{jms_193_33} & 505 & -- & 375 & 631.5437158(56)\cite{jms_219_191} & 628 &-- & 491\\
 $\tau_{\ce{^xH 1CO}}$ &1010.8\cite{cjs_18_135} & 1029 & -- & 1013 & 870\cite{jcp_27_1305}& 872 &-- & 861\\
 \ce{OCO-CO^xH 2} def. &558.2722502(62)\cite{jms_193_33}& 556 & -- &622 & 620.5683857(57)\cite{jms_219_191} & 617 &-- & 652\\
$\delta_{\ce{CO^xH 2}}$ &972.8520(1)\cite{jms_198_110,qe_29_226}& 969 &904\cite{pccp_22_25492}& 905 &1297\cite{jcp_85_4779}& 1294 &-- & 1235 \\
$\gamma_{\ce{C^xH 1}}$ &1366.48430(39)\cite{jms_334_22} & 1362 & --& 1383 & 970\cite{jcp_85_4779}& 971 &-- & 986\\
$\nu_{\ce{C-O}}$ &1177.09378(2) \cite{jms_191_343} & 1179 & -- &1162  & 1142.31075(2)\cite{jms_190_125}&  1139 & --& 1130\\
$\nu_{\ce{C=O}}$ &1772.119(24)\cite{jms_213_1} & 1770 & 1819\cite{pccp_22_25492} & 1815 & 1725.874974(34)\cite{sapa_55_2601} & 1723 & 1790\cite{pccp_22_25492}& 1781\\ 
$\nu_{\ce{C^xH 1}}$ &2938-2942\cite{jcp_85_4779} &2934  & 2871\cite{pccp_22_25492} & 2869 & 2219.6896(2)\cite{jms_198_387} & 2216 & 2167\cite{pccp_22_25492}& 2167\\ 
$\nu_{\ce{O^xH 2}}$ &2631.63835(17)\cite{jms_349_43} & 2629  & 2685\cite{pccp_22_25492} & 2683 & 3566\cite{jcp_85_4779} & 3568 &3635\cite{pccp_22_25492}& 3625\\
\hline
\end{tabular}}

\footnotetext[1]{From Richter and Carbonni\`ere.\cite{jcp_6_064303}}
\footnotetext[2]{From Aerts \textit{et al.}\cite{jcp_152_024305}}
\end{center}
\label{table:fundaformic}
\end{table}%

\section{Computational details}
\subsection{Multi Configurational Time Dependent Hartree method}

The fundamental problem faced in molecular quantum mechanical studies
is the (extremely) large dimension of the primitive basis set which is a product
basis built from 1D basis functions for each nuclear degree of freedom. To
reduce the size of the quantum mechanical problem, several strategies
have been developed to work in a smaller subspace, or ``active'' space
of the primitive space.
The MCTDH method \cite{mey90:73,man92:3199,bec00:1,mey03:251}
uses a time-dependent active space,  built by time--dependent
functions $\varphi(q,t)$, called single--particle functions
(SPFs). The MCTDH active space is thus the direct product space of the SPFs.
It is worth noting that the SPFs may be multi-dimensional functions and the
coordinate $q$ may be a collective one, where $q=(Q_k,\cdots,Q_l)$ and $Q_i$ is a 1D coordinate.
As the SPFs are time-dependent, they follow the wavepacket and adapt along the propagation such that
often a rather small number of SPFs suffices for convergence.

The MCTDH wavefunction reads
\begin{eqnarray}
\Psi(Q_1,\cdots,Q_f,t) & \equiv & \Psi(q_1,\cdots,q_p,t) \nonumber \\
& = & \sum_{j_1}^{n_1} \cdots \sum_{j_p}^{n_p} A_{j_1,\cdots,j_p} \,
\prod_{\kappa=1}^{p} \varphi^{(\kappa)}_{j_\kappa}(q_\kappa,t) \; ,
\end{eqnarray}
where $f$ denotes the number of nuclear degrees of freedom, $p$ the number
of MCTDH particles, also called combined modes, and there are
$n_\kappa$ SPFs for the $\kappa$'s particle. 
In the present computations, the mode combinations and number of SPFs varied from one simulation to the other. All details regarding the discrete variable representation (DVR) and SPF bases are given in the Supplementary Material.
The equations of motion \cite{mey90:73,man92:3199,bec00:1,mey03:251}
for the coefficient vector $A$ and for the SPFs are derived from the Dirack-Frenkel
variational principle. Thus, MCTDH uses
-- in a variational sense -- optimal SPFs, as this ensures
optimal fast convergence.

\subsection{Improved Relaxation}
\label{sec:improvedrelax}
The improved relaxation method\cite{mey06:179,dor08:224109} implemented in the MCTDH package was used to compute all vibrational states in this work. 
The eigenstates of the Hamiltonian are obtained in an iterative fashion. The equations  
can be derived via a standard {\it time-independent}
variational principle
$\delta\big\{\big<\Psi\big|H\big|\Psi\big> - \text{constraints}\big\} = 0$, \textit{i.e.},
\begin{equation}
   \delta\big\{\big<\Psi\big|H\big|\Psi\big> - E\big(\sum_J A^*_J A_J-1\big)-
    \sum_{\kappa=1}^f\sum_{j,l=1}^{n_\kappa}\epsilon_{jl}^{(\kappa)}
    \big(\big<\varphi_j^{(\kappa)}|\varphi_l^{(\kappa)}\big>-\delta_{jl}\big)\big\} = 0 \,.
\end{equation}
This approach is thus similar to the MCSCF method in quantum chemistry.
Therefore,
\begin{equation}
    \sum_L \langle\Phi_J|\hat{H}|\Phi_L\rangle A_L = E\phantom{.}A_J,
    \label{eq:Avector}
\end{equation}
where $\Phi$ are the Hartree product of SPFs. For the SPFs,
\begin{equation}
    \sum_{l=1}^{n_\kappa}\langle\mathbf{H}\rangle^{(\kappa)}_{jl}\varphi_l^{(\kappa)}=\sum_{l=1}^{n_\kappa}\epsilon_{jl}^{(\kappa)}\varphi_{l}^{(\kappa)}
\end{equation}
which can be reformulated as a propagation in imaginary time (\(\tau=-it\)):
\begin{equation}
    \frac{\partial}{\partial \tau}\varphi_{j}^{(\kappa)}:=-(1-P^{(\kappa)})\sum_{k,l=1}^{n_\kappa}(\rho^{(\kappa)})^{-1}_{jk}\langle\hat{H}\rangle^{(\kappa)}_{kl}\varphi_{j}^{(\kappa)}=0.
    \label{eq:imagpropa}
\end{equation}

An iterative algorithm starts from an initial guess state and solves Eq. \ref{eq:Avector} within the SPF basis functions using a Davidson diagonalisation. The coefficients are chosen corresponding to the state that is closest to the defined target. Therefore, it requires a  guess of the wavefunction which should have a reasonable overlap with the desired eigenstate. The SPFs are relaxed (Eq. \ref{eq:imagpropa}) with fixed coefficients, and the procedure is repeated with the new basis until convergence.

\subsection{Potential energy surface}
Richter and Carbonni\`ere\cite{jcp_6_064303} reported a full dimensional ground state potential energy surface for formic acid fit using the iterative approach of AGAPES\cite{jcp_136_224105} from 660 ab initio energies at the CCSD(T)-F12a/aug-cc-pVTZ level of theory. The resulting potential is in sum of product form and uses internal valence coordinates for direct use in MCTDH. The expansion of the potential is in ``High Dimensional Model Representation" form, with each term of the expansion successively improved until convergence. The multidimensional terms can be interpreted as coupling terms for the coordinate combinations, and were found to quickly converge to zero with increasing order.
The surface covers both \textit{trans} and \textit{cis} minima and is valid up to 6000 cm\textsuperscript{-1} above the \textit{trans}-HCOOH ZPE. It encompasses well the \textit{trans} to \textit{cis} isomerisation path as the barrier is at about 4083 cm\textsuperscript{-1} above the \textit{trans}-HCOOH ZPE. 

Although some mode couplings and effects could be determined from the analysis of the potential expansion through its raw numerical terms, the intramolecular dynamics at energies around the isomerisation barrier is expected to be non-trivial. Also, effects from the kinetic energy operator could be significant and such predictions about mode coupling can be more challenging to make simply from the analysis of the PES expansion coefficients. 

\subsection{Local modes}
\label{sec:localmodes}
Given the expected large motion in the torsion coordinate induced by the \textit{trans}-\textit{cis} isomerization, the choice of coordinates of Richter and Carbonni\`ere are particularly appropriate in this case. The internal valence coordinates (see Fig. \ref{fig:zmatrix}) minimize potential energy couplings in the local mode region, and therefore, they maximize the separation of vibrational motion to some extent. For instance, the torsional motion involved in the \textit{trans}-\textit{cis} isomerization is mainly characterised by the $\tau_2$ coordinate. 

To study the energy flow driven by intramolecular couplings in formic acid when a local mode is excited by \textit{n} quanta, we define the energy in the (local) mode $i$ as
\begin{equation}
E_i(t)=\langle \Psi(t)|\hat{h}_i|\Psi(t)\rangle,
\end{equation}
where $\Psi(t)$ is the time-dependent wavefunction and $\hat{h}_i$ is the Hamiltonian operator associated with the curvilinear local mode $i$. The corresponding 9 Hamiltonian operators are obtained from the complete Hamiltonian with all inactive coordinates frozen to their values at the \textit{trans} minimum. 
\par To generate the local torsional excitations, 8-dimensional Hamiltonians were retrieved from the full dimensional potential by removing all differential operators acting on $\tau_2$, and for all remaining terms, $\tau_2$ was set to its equilibrium value, \textit{i.e.}, 0 for the \textit{trans}  isomer. The wavefunctions are constructed from the product of the eigenfunction of the local mode operator $\hat{h}_{\tau_2}$ and the ground state of the 8 dimensional Hamiltonian obtained with the improved relaxation method.\cite{mey03:251,mey05:66,dor08:224109,mey09:book,mey12:351} The local vibrational levels will be further identified with $v_i$, the vibrational quantum number associated with the respective mode of vibration. Naturally, we have the stretching modes associated with the bond distance coordinates, $\theta_1$ with the \ce{OCO-CO^xH 2} deformation, $\theta_2$ with the $\gamma_{\ce{C^xH 1}}$ rocking, $\theta_3$ with the $\delta_{\ce{CO^xH 2}}$ bending, $\tau_1$ with the 
$\tau_{\ce{^xH 1CO}}$ torsion, and $\tau_2$ with the 
$\tau_{\ce{CO^xH 2}}$ torsion. 
The features of the dynamics will be discussed based on fractional energies defined by

\begin{equation}
F_i^n(t)=\frac{E_i(t)-E_i(t=0)}{E_{\tau_2}^n-E_{\tau_2}^0}
\end{equation}
 except for the torsional coordinate $\tau_2$ for which: 
 \begin{equation}
 F_{\tau_2}^n(t)=\frac{E_{\tau_2}(t)-E_{\tau_2}^0}{E_{\tau_2}^n-E_{\tau_2}^0}
 \end{equation}
 where $E_{\tau_2}^n$ is the initial energy of the $n$th excited eigenstate of $\hat{h}_{\tau_2}$ and $E_{\tau_2}^0$ its zero point energy ($n=0$). An analogous procedure was also used to produce local excitations in the $r_1$ coordinate (${\ce{C-O}}$ stretch) and study the subsequent dynamics driven by the IVR. By definition, the wavefunctions generated with the 8 dimensional Hamiltonians do not include couplings with the excited mode. Combined with the approximate nature of the local mode operators which do not include couplings by construction, the fraction of energy in the initially excited mode exceeds 100 \% in the first moments of the simulations.  

Additionally, the isomerization dynamics will be described by the probability to find the molecule in the initial well:
\begin{equation}
P_{\text{surv}}(t)=\langle \Psi(t)|\Theta_{\tau_2}|\Psi(t)\rangle
\end{equation}
where
\begin{equation}
\Theta_{\tau_2}=
\begin{cases}
1 \phantom{--}\forall \phantom{|}\tau_2 \in \text{initial well}\\
0 \phantom{--}\forall \phantom{|}\tau_2 \notin \text{initial well.}
\end{cases}
\end{equation}

Initial wells are defined by the intervals of the torsional angle $\tau_2$ coordinate, $\tau_2\in[-\pi/2,\pi/2]$ and $\tau_2\in[\pi/2,3\pi/2]$ for the \textit{trans} and \textit{cis} isomers, respectively.

\section{Results and discussion}

\begin{table}
 \caption{Vibrational energies of local ($E_{\nu_{\tau_2}}$) and full-dimensional ($E_{\tau_{\ce{CO^xH 2}}}$) torsional $\tau_2$  overtones in \textit{trans}-formic acid and deuterated forms with overlap of the (full-dimensional) eigenstate wavefunction with its zero-order counterpart ($\langle \Psi^{0}|\Psi\rangle$) by increasing number of quanta ($v_{\tau_2}$).
    All energies, including the isomerisation barriers, are given with respect to the isotopologue vibrational zero point energy in cm\textsuperscript{-1}.
    Values in brackets are tentative mode assignments.}
    \centering
    \begin{tabular}{|c|ccc|ccc|ccc|ccc|}
    \hline
    \hline
 & \multicolumn{3}{c|}{HCOOH}  & \multicolumn{3}{c|}{DCOOD}  & \multicolumn{3}{c|}{DCOOH}  & \multicolumn{3}{c|}{HCOOD} \\
 \hline
 $\phantom{.}v_{\tau_2}$& $E_{\nu_{\tau_2}}$ & $E_{\tau_{\ce{CO^xH 2}}}$\footnotemark[1] & $\langle \Psi^{0}|\Psi\rangle$ & $E_{\nu_{\tau_2}}$ & $E_{\tau_{\ce{CO^xH 2}}}$\footnotemark[2] & $\langle \Psi^{0}|\Psi\rangle$ & $E_{\nu_{\tau_2}}$ & $E_{\tau_{\ce{CO^xH 2}}}$\footnotemark[2] & $\langle \Psi^{0}|\Psi\rangle$ & $E_{\nu_{\tau_2}}$ & $E_{\tau_{\ce{CO^xH 2}}}$\footnotemark[2] & $\langle \Psi^{0}|\Psi\rangle$ \\
 \hline
2 & 1276 & 1216 & 0.61 & 1045 & 960 & 0.71 & 1295 & 1202 & 0.63 & 1038 & 1006 & 0.56 \\
3 & 1880 & 1796 & 0.60 & 1543 & 1406 & 0.71 & 1898 & 1750 & 0.43 & 1531 & 1441 & 0.59 \\
4 & 2451 & (2332) & 0.28 & 2034 & 1843 & 0.71 & 2470 & 2282 & 0.51 & 2005 & 1890 & 0.57 \\
5 & 2987 & 2799 & 0.51 & 2461 & 2263 & 0.67 & 3005 & 2775 & 0.58 & 2454 & 2322 & 0.57 \\
6 & 3508 & 3264\footnotemark[3] & 0.55 & 2891 & (3055) & 0.27 & 3527 &  &  & 2884 & (2733)\footnotemark[3] & 0.23 \\
7 &  &  &  & 3297 &  &  &  &  &  & 3290 & (3135)\footnotemark[3] & 0.32 \\
8 &  &  &  & 3704 &  &  &  &  &  & 3696 &&\\
9 &  &  &  & 4147 &  &  &  &  &  & 4088&& \\
10 &  &  &  & 4497 &&&&&&&&\\
\hline
Barrier & \multicolumn{3}{c|}{4053}  & \multicolumn{3}{c|}{4179}  & \multicolumn{3}{c|}{4100}   & \multicolumn{3}{c|}{4171} \\
\hline
\hline
    \end{tabular}

    \footnotetext[1]{From Richter and Carbonni\`ere.\cite{jcp_6_064303}}
    \footnotetext[2]{From Aerts \textit{et al.}\cite{jcp_152_024305}}
    \footnotetext[3]{This work.}

    \label{table:local_full_overlap}
\end{table}

From locally excited torsion overtones (see Section~\ref{sec:localmodes}), wavepackets are propagated on the full-dimensional potential energy surface in valence coordinates. We report the energies above the vibrational ZPE of the local torsion modes $\tau_2$ of formic acid and deuterated forms in the \textit{trans} well in Table~\ref{table:local_full_overlap}. For comparison, we also report the eigenenergies of the torsional vibrational modes obtained with the full-dimensional Hamiltonian, \textit{i.e.}, that includes all couplings. Eigenenergies given in brackets in Table~\ref{table:local_full_overlap} correspond to eigenstates with the largest overlap between the wavefunction and its zero-order (local) counterpart. However, in these cases, the overlap is smaller than 0.5 and therefore the assignment should be taken as tentative. These states are attributed to a specific zero-order label (number of quanta $v_{\tau_2}$) but this label loses its meaning due to couplings. 
Discrepancies between local modes energies and eigenenergies appear due to the absence of coupling between the excited vibrational mode and others, especially for high energy states. The lack of couplings also leads to some fractions of energy exceeding 100\% during the propagations. We computed locally excited vibrational states within the valid energy window of the PES, and stopped increasing the number of excitation quanta once isomerization was observed.

The difficulty to converge eigenstates high in energy in full-dimensionality increases due to mixing with coupled modes, which is quantified in Table \ref{table:local_full_overlap} by the overlap of the converged wavefunction of eigenstates including all mode-couplings with their zero-order counterpart. The latter is constructed by the Hartree product of all (vibrational) mode-Hamiltonians eigenfunctions. In practice, the zero-order wavefunctions are converged using the improved relaxation method with mode-couplings removed from the Hamiltonian. The zero-order wavefunction essentially being the initial guess of the procedure, it usually becomes increasingly difficult to converge the states going higher in energy, due to the fact that an accurate description of excited states requires an increasing number of SPFs. Unambiguous assignment of the converged eigenstates, by inspection of the reduced densities or by calculation of the overlap with the zero-order states wavefunction, is not guaranteed especially high in energy when the density of states becomes too significant for the approach to be sensible in the context of this study: this is the reason why some values are given in parentheses in Table \ref{table:local_full_overlap}.

\subsection{\textit{trans}-\ce{HCOOH} $\tau_2$ torsion}
\begin{figure}[htbp]
   \centering
   \includegraphics[width = \columnwidth]{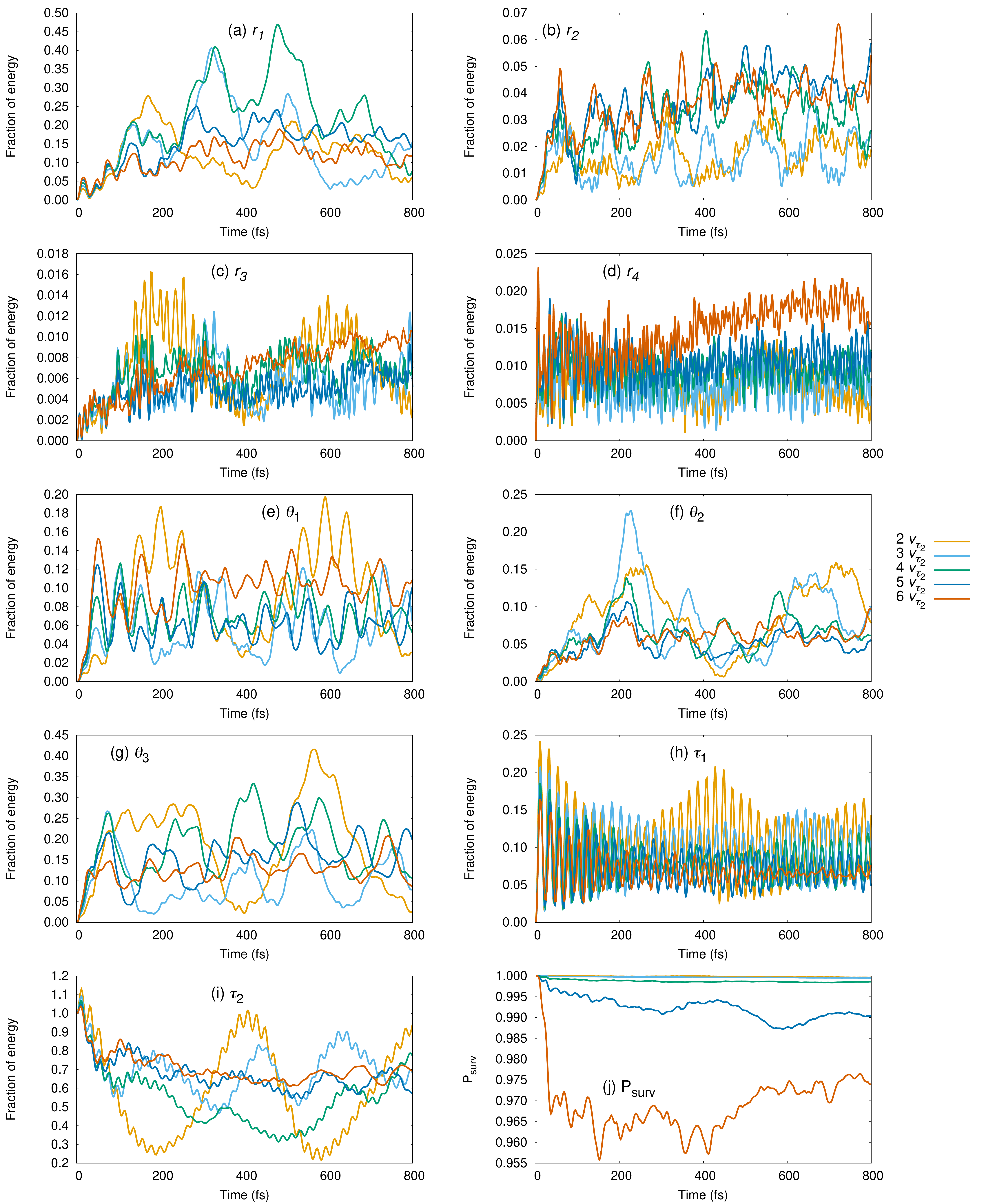} 

\caption{Fraction of energy in all nine modes ({a} to {i}) and survival probability $P_{\text{surv}}$ (j) of the propagations of overtones of local torsional mode $\tau_{2}$ of \textit{trans}-HCOOH with 2 to 6 quanta. Note the change of scales for the y-axis for (a)-(j).}
\label{fig:energyflow_trans_hcooh}
\end{figure}

The energy flows driven by the IVR, as well as $P_\text{surv}$, which quantifies the probability to find the wavefunction in the initial (\textit{trans}) well are shown in Fig. \ref{fig:energyflow_trans_hcooh} for the local mode excitations with different numbers of quanta in the torsional mode. The flows of energy demonstrate that the torsional angle coordinate is not isolated. Additionally, it is clear that the mode couplings have different magnitudes and will lead to non-statistical IVR dynamics which is also in agreement with experimental observations.\cite{jcp_154_064301} The prominent feature of the energy flows is the well known\cite{jcp_126_164305,jmst_795_54,jcp_76_886,aipa_1_015021} strong Fermi resonance between the 2 $\tau_{\text{COH2}}$ ($2\tau_2$) out-of-the-plane overtone with the $\delta_\text{COH}$ ($\theta_3$) bending mode leading to a strong and reversible energy flow between the states, see Figs. \ref{fig:energyflow_trans_hcooh} (g) and \ref{fig:energyflow_trans_hcooh} (i). The pattern of energy flow remains distinguished between the two modes with an increased number of quanta, which corresponds to coupled states that share the polyad number $N_p=v_{\theta_3}+v_{\tau_2}/2$ that was previously identified.\cite{aipa_1_015021,jpca_120_9815,jcp_6_064303,jcp_154_064301}
A significant part of energy flows from the excited mode $\tau_2$ through the other out-of-the-plane mode $\tau_1$ and acts as an energy reservoir from the torsional mode, see Fig.~\ref{fig:energyflow_trans_hcooh} (h). It confirms a resonance between the two modes. 


Interestingly, starting at $v_{\tau_2}=3$, the coupling with the $\nu_\text{\ce{C-O}}$ stretching mode becomes more significant, see Fig. \ref{fig:energyflow_trans_hcooh} (a). 
Also, the (modest) coupling with the $\nu_\text{\ce{C=O}}$ stretching mode ($r_2$) must be pointed out, see Fig. \ref{fig:energyflow_trans_hcooh} (b), which is in good agreement with the assignments of Perrin \textit{et al.} \cite{jqsrt_110_743} of the interacting states $\nu_\text{\ce{C=O}}, \nu_\text{\ce{C-O}} + \ce{OCO}-\ce{COH 2} \text{ def.}, \nu_\text{\ce{C-O}} + \tau_{\ce{COH 2}}, 3\tau_{\ce{COH 2}}$ and $\ce{OCO}-\ce{COH 2} \text{ def.}+2\tau_{\ce{COH 2}}$. We also observe a strong coupling with the dark $\gamma_{\ce{CH 1}}$ rocking mode ($\theta_2$) for $v_{\tau_2}=3$ leading to periodic energy flows, see Fig. \ref{fig:energyflow_trans_hcooh} (f). The coupling of the torsional mode with the stretching modes is strongest for the $\ce{C-O}$ ($r_1$) stretch and an order of magnitude smaller for the $\ce{C=O}$ ($r_2$), $\ce{C-H 1}$ ($r_3$), and \ce{O-H 2} ($r_4$) stretches.

Increasing the initial number of quanta in the out-of-the plane mode $\tau_2$, we observe a quasi systematic increase of energy flowing to the $\nu_\text{\ce{C=O}}$ and $\nu_\text{\ce{C-H 1}}$ bright stretching modes; see Figs~\ref{fig:energyflow_trans_hcooh} (b) and \ref{fig:energyflow_trans_hcooh} (c), however, it remains small as less than 7\% and 2\% of the initial energy flows towards these modes respectively. Comparatively, the couplings with dark modes lead to energy flows that are approximatively 2 to 5 times more efficient. On the other hand, the coupling with the $\nu_\text{\ce{C-O}}$ stretching mode is still significant but does not exhibit systematic changes with increasing quanta in the torsional coordinate. Reaching $v_{\tau_2}=6$, the quasi-periodic energy flows disappear between the initially excited torsion angle mode and the other modes. The isomerisation mechanism is initiated and part (\textit{ca.} 0.5\%) of the wavepacket is transfered from the \textit{trans} to the \textit{cis} well within 100 fs. The transition occurs at an energy well below the isomerisation barrier of 4083 cm\textsuperscript{-1}, \textit{i.e.}, with around 3508 cm\textsuperscript{-1} in the torsion coordinate (both with respect to the trans vibrational ZPE, see Table \ref{table:local_full_overlap}) which indicates a tunnelling effect. No strong periodic feature is observed in the evolution of $P_{surv}$ on the timescale studied, revealing an irreversible intramolecular process.

With further increase of the energy in the torsion angle coordinate $\tau_2$, new couplings tend to suppress the structure of the energy flow in the other modes which becomes somewhat non-specific and (mostly) irreversible as the density of states increases. Only the $\nu_\text{\ce{C-O}}$ stretch shows a significant coupling with $\tau_2$ amongst the bright modes. This is an interesting result since this could be exploited for laser-induced isomerisation as this stretching mode possesses a strong oscillator strength.\cite{jgra_104_18661} 

For completeness, we have optimized the eigenstates that are in interaction with the 6$\tau_{\ce{COH 2}}$ eigenstate; their energies and overlaps with zero-order counterparts are given in Table~\ref{tab:6v9_interacting_states}. Again, we point out the effect of mode couplings which make the overlaps of the eigenstates with their zero-order counterparts rather low. Although these are particularly low for the combination modes reported, \textit{i.e.}, 4$\tau_{\ce{COH 2}}$+\ce{OCO}-\ce{COH 2} def., 4$\tau_{\ce{COH 2}}$+$\delta_{\ce{COH 2}}$, and 4$\tau_{\ce{COH 2}}$+$\nu_{\ce{CO}}$, their assignment by analysis of reduced densities along the coordinates is unambiguous. The 6$\tau_{\ce{COH 2}}$ eigenstate has largest overlap with its zero-order counterpart (56\%), as well as the 4$\tau_2$+$\theta_1$ (10\%), and 4$\tau_2$+$\theta_3$ (8\%) combination modes. This is in agreement with the dynamics described in Fig. \ref{fig:energyflow_trans_hcooh} as, in the first moments, the largest fraction of energy flows into the \ce{OCO}-\ce{COH 2} def. mode (see panel (e)), and the $\delta_{\ce{COH 2}}$ mode (see panel (g)). From Table~\ref{tab:6v9_interacting_states}, we also see that the largest overlap of the 4$\tau_{\ce{COH 2}}$+\ce{OCO}-\ce{COH 2} def. and 4$\tau_{\ce{COH 2}}$+$\delta_{\ce{COH 2}}$ eigenstates, besides their zero-order counterpart, is with the 4$\tau_2$+$r_1$ mode. That observation supports the indirect coupling that we identified between the out-of-the plane $\tau_{\ce{COH 2}}$ torsion and the $\nu_{\ce{C-O}}$ stretch.

\begin{table}[]
    \centering
        \caption{Eigenstate energies and overlap of vibrational states of \textit{trans}-\ce{HCOOH} that interact with 6$\tau_{\ce{COH}}$ and their zero-order counterparts.}
    \begin{tabular}{|cc|cccc|}
    \hline
    \hline
        \multirow{2}{*}{Assignment} & \multirow{2}{*}{Energy (cm\textsuperscript{-1})} &\multicolumn{4}{|c|}{$\langle \Psi^{0}|\Psi\rangle$}  \\
        &&6$\tau_2$&4$\tau_2$+$\theta_1$&4$\tau_2$+$\theta_3$&4$\tau_2$+$r_1$\\
        \hline
         6$\tau_{\ce{CO^xH 2}}$& 3264&0.56&0.10&0.08&0.01\\
         4$\tau_{\ce{CO^xH 2}}$+\ce{OCO-CO^xH 2} def.&2953&0.03&0.15&0.001&0.09\\
         4$\tau_{\ce{CO^xH 2 2}}$+$\delta_{\ce{CO^xH2}}$&3606&0.03&0.007&0.15&0.08\\
         4$\tau_{\ce{CO^xH 2}}$+$\nu_{\ce{C-O}}$&3531&0.004&0.02&0.04&0.16\\
         \hline
         \hline
    \end{tabular}
    \label{tab:6v9_interacting_states}
\end{table}

\subsection{\textit{trans}-\ce{HCOOH} $r_1$ stretch}

As pointed out in the previous section, a significant coupling exists between the torsion coordinate that characterises the isomerisation of formic acid and other vibrational modes of formic acid. Since the \ce{C-O} stretch is IR active, its excitation could favour the isomerisation process, by the specific flow of energy towards the torsion coordinate $\tau_2$, but also weakens the delocalised $\pi$ electrons along the \ce{O=C-O} chain that imposes the planarity in formic acid.	 

\begin{figure}
 \includegraphics[width = \columnwidth]{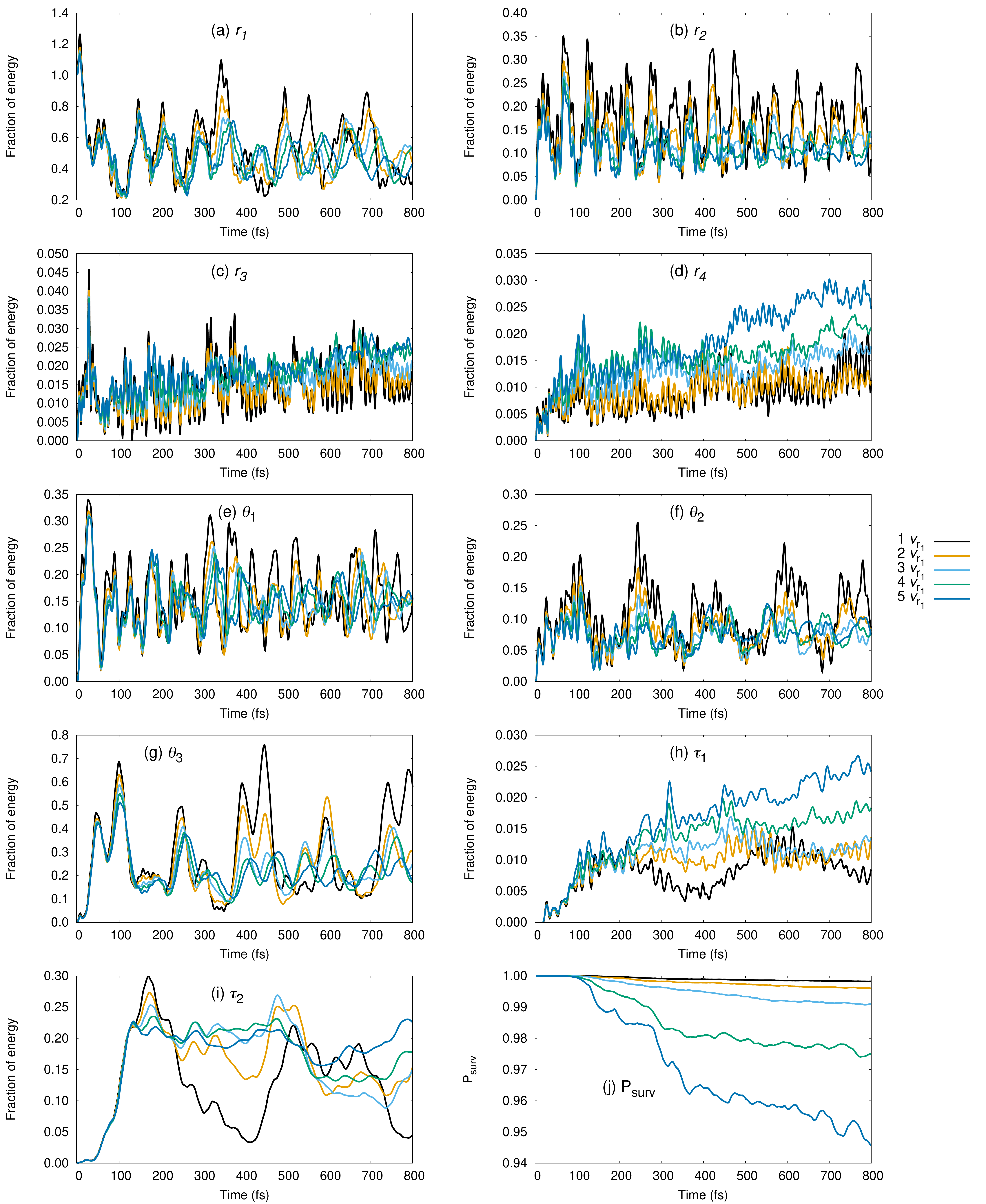} 

\caption{Fraction of energy in all nine modes ({a} to {i}) and survival probability $P_{\text{surv}}$ (j) of the propagations of the fundamental and overtones of local $\ce{C-O}$ stretch mode of \textit{trans}-HCOOH with 1 to 5 quanta. Note the change of scales for the y-axis for (a)-(j).}
\label{fig:energyflow_CO_trans_hcooh}
\end{figure}

We report the local ${\ce{C-O}}$ stretch vibrational frequencies of \textit{trans}-\ce{HCOOH} in Table \ref{table:co-trans-HCOOH} with respect to the vibrational ZPE. We also include energies in full-dimension for states that could be identified given the increased mixing high in energy that is quantified by the overlap of the converged wavefunction with its zero-order counterpart, also in Table \ref{table:co-trans-HCOOH}. The values underline the difficulty to converge the eigenstates and make unambiguous assignments high in energy; this assignment is of course not applicable to ``local" states. In Fig.~\ref{fig:energyflow_CO_trans_hcooh}, we show the energy flows from the excitations towards all other coordinates along with $P_\text{surv}$ which quantifies the probability to find the wavefunction in the initial (\textit{trans}) well. The energy flows are well structured up to about 200 fs in all modes and for all excitations. A prominent exchange takes place between the initially excited mode and the $\delta_{\ce{COH 2}}$ bending mode through their strong coupling, see Fig. \ref{fig:energyflow_CO_trans_hcooh} (g). Fast exchanges of energy with large amplitudes remain mostly periodic between the $\nu_{\ce{C-O}}$ stretch and the $\nu_{\ce{C=O}}$ stretch, $\gamma_{\ce{CH 1}}$ rocking and \ce{OCO}-\ce{COH 2} deformation, although the amplitude decreases with increasing excitation energy, see Figs. \ref{fig:energyflow_CO_trans_hcooh} (a), (b) and \ref{fig:energyflow_CO_trans_hcooh} (f) and (e), respectively.

\begin{table}
\caption{Vibrational energies of local ($E_{\nu_{r_1}}$) and full-dimensional ($E_{\nu_{\ce{C-O}}}$) $\nu_{\ce{C-O}}$ stretch fundamental and overtones in \textit{trans}-HCOOH with overlap of the (full-dimensional) eigenstate wavefunction with its zero-order counterpart ($\langle\Psi^0|\Psi\rangle$) by increasing number of excitation quanta ($v_{r_{1}}$). All energies are given with respect to the vibrational zero point energy in cm\textsuperscript{-1}.  Values in brackets are tentative mode assignments.}
\begin{center}
\begin{tabular}{|c|ccc|}
\hline
\hline
&\multicolumn{3}{c|}{HCOOH}\\
\hline
$v_{r_{1}}$ & $E_{\nu_{r_1}}$&$E_{\nu_{\ce{C-O}}}$\footnote{From Richter and Carbonni\`ere.\cite{jcp_6_064303}}&  $\langle\Psi^0|\Psi\rangle$\\

\hline
1 & 1239&1301&0.39\\
2&2456&(2498)&0.15\\
3&3652&(3482)\footnote{This work.}&0.07\\
4&4827&&\\
5&5982&&\\
\hline
\hline
\end{tabular}
\end{center}

\label{table:co-trans-HCOOH}
\end{table}%

\begin{figure}
\includegraphics[width =0.4 \columnwidth]{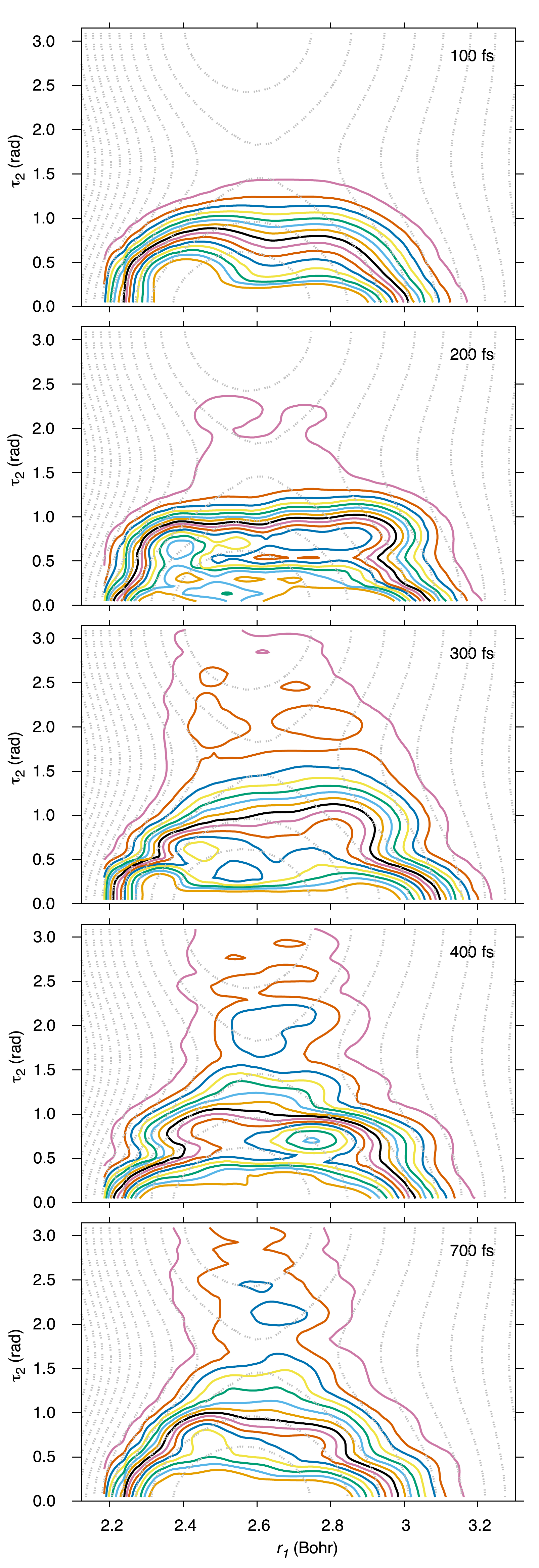} 

\caption{Reduced density contours of the excited wavepacket with initially 5 quanta in the local $\nu_{\ce{C-O}}$ stretch mode of \textit{trans}-\ce{HCOOH}. The equidensity lines are logarithmically spaced (continuous) and the potential energy cut is overlayed (dashed) with all other coordinates fixed at their values at the \textit{trans} equilibrium geometry.}
\label{fig:densityWavePacket}
\end{figure}

Energy exchange with the $\delta_{\ce{COH 2}}$ mode happens on a longer timescale and is somewhat complementary with the flow of energy in the torsion angle coordinate. This complementarity indicates an efficient indirect coupling between the $\nu_{\ce{C-O}}$ stretch and the $\tau_{\text{COH 2}}$ torsion, where the fraction of energy flowing to the torsional coordinate greatly increases during the first 200 fs of the dynamics. Above $v_{r_1}=3$, the amplitude of the energy flows in the $\delta_{\ce{COH 2}}$ mode decreases dramatically after 300 fs, see Fig.~\ref{fig:energyflow_CO_trans_hcooh} (g). This corresponds to a regime during which a steady part of the energy remains in the torsion angle mode and isomerisation takes place. Fig.~\ref{fig:densityWavePacket} shows that the wavepacket (here with $\nu_{r_{1}}=5$) starts to migrate towards the \textit{cis} isomer geometry after 100 fs with a build-up of its reduced density over time.

At 800 fs, a significant part of the energy, greater with increasing excitation energy, remains in the torsion angle mode coordinate without significant apparent periodic flow above 3 quanta. These observations support our assumption that the strong oscillator strength of the $\nu_{\ce{C-O}}$ stretch\cite{jgra_104_18661} and its efficient indirect coupling with the $\tau_{\text{COH 2}}$ torsion mode could be exploited for laser-induced isomerisation in \textit{trans}-HCOOH. Since the coupling exists, and is exhibited in the IVR no matter which local state ($\tau_2$ torsion or \ce{C-O} stretch) is initially populated, only the initial population of torsional modes is considered for the other isotopologues.

\subsection{\textit{trans}-\ce{DCOOD} $\tau_2$ torsion}

\begin{figure}
 \includegraphics[width = \columnwidth]{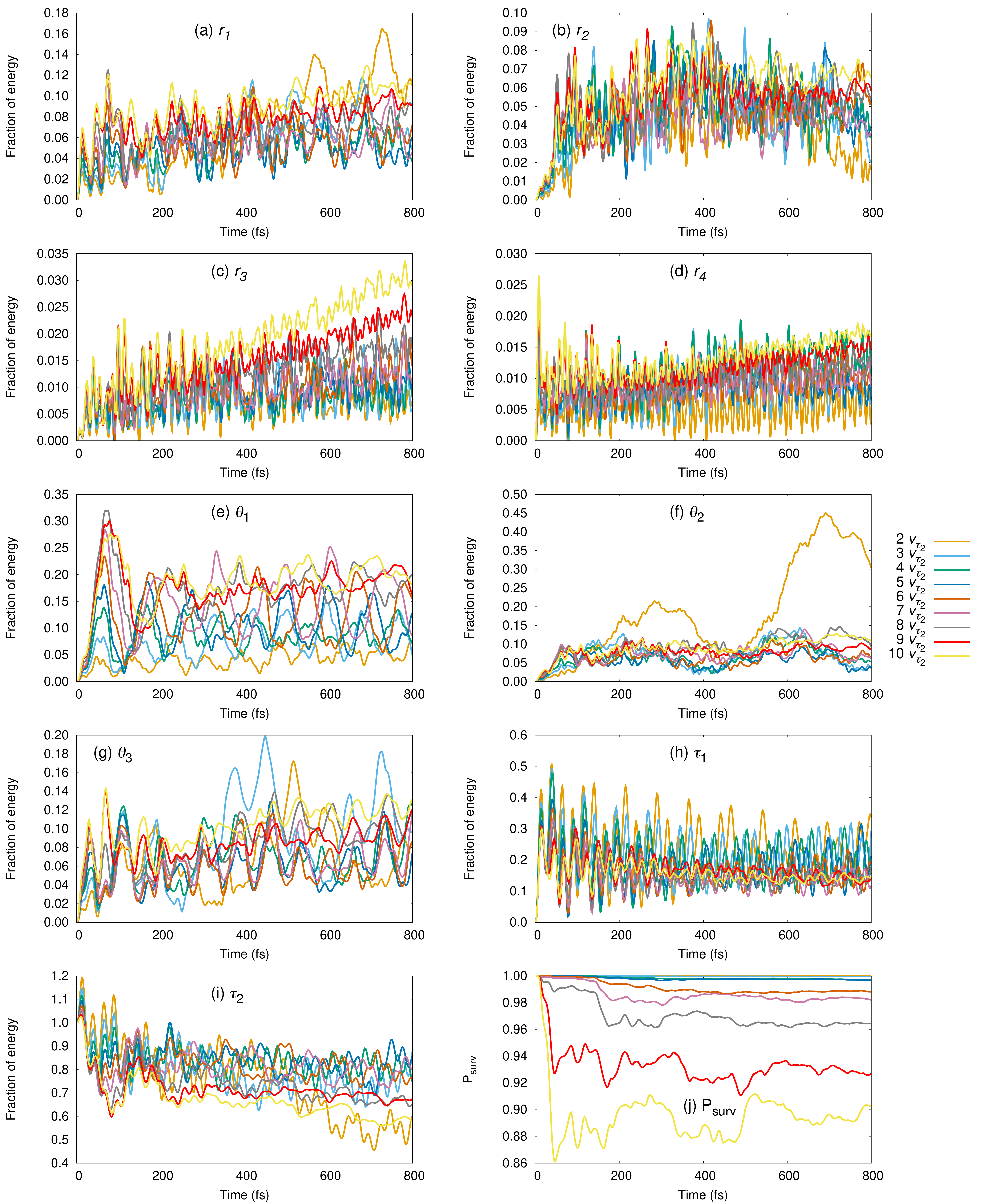} 
\caption{Fraction of energy in all nine modes ({a} to {i}) and survival probability $P_{\text{surv}}$ (j) of the propagations of overtones of local torsional mode $\tau_{2}$ of \textit{trans}-DCOOD with 2 to 10 quanta. Note the change of scales for the y-axis for (a)-(j).}
\label{fig:energyflow_torsion_trans_DcooD}
\end{figure}

The IVR dynamics of the doubly deuterated form of formic acid $\ce{DCOOD}$ is very different from the un-deuterated case as can be seen in Fig. \ref{fig:energyflow_torsion_trans_DcooD} which depicts the fraction of energy that flows through the nine modes of formic acid from the overtones of the local torsional mode $\tau_2$ with excitation quanta from 2 to 10. We can achieve higher relative energies and number of excitation quanta in the case of \ce{DCOOD} within the range of accuracy of the potential because the ZPE of \ce{DCOOD} is lower than the one of HCOOH.

Although interesting mode couplings were observed in the previous section, the isotopic substitution does not seem to induce new resonances with the torsional mode of interest $\tau_{\ce{COD 2}}$. We point out the isolated nature of both the $\nu_{\ce{C-D 1}}$ and $\nu_{\ce{O-D 2}}$ stretches within the excitation energies considered, see Figs~\ref{fig:energyflow_torsion_trans_DcooD} (c) and \ref{fig:energyflow_torsion_trans_DcooD} (d). Isotopic substitution even destroys the coupling with the $\nu_{\ce{C-O}}$ stretch, see Fig~\ref{fig:energyflow_torsion_trans_DcooD} (a), and therefore the energy deposited in the local torsional mode $\tau_2$ is found to evolve in an almost statistical manner; the energy flows through all modes without apparent strong coupling with one of them. An exception is found for the first overtone of the torsional mode ($v_{\tau_2}=2$) arising probably from an accidental resonance with the $\gamma_{\ce{CD 1}}$ rocking mode as shown in Fig. \ref{fig:energyflow_torsion_trans_DcooD} (f).

A substantial isomerisation is observed for 9 and 10 quanta but this is not surprising as their vibrational energies are at or above the energy barrier to isomerisation, see Table \ref{table:local_full_overlap}. Again, the isomerisation process already takes place below the barrier of about 4179 cm\textsuperscript{-1} above the \textit{trans}-DCOOD ZPE. However, no efficient transfer towards bright vibrational mode(s) appears during the timescale of our simulation starting from local overtones of the local torsional mode $\tau_2$. Therefore, laser-induced isomerisation of \textit{trans}-DCOOD would certainly not be achieved taking advantage of vibrational couplings as it was suggested for \textit{trans}-HCOOH in the previous section.

\subsection{\textit{trans}-\ce{DCOOH} $\tau_2$ torsion}

\begin{figure}
 \includegraphics[width = \columnwidth]{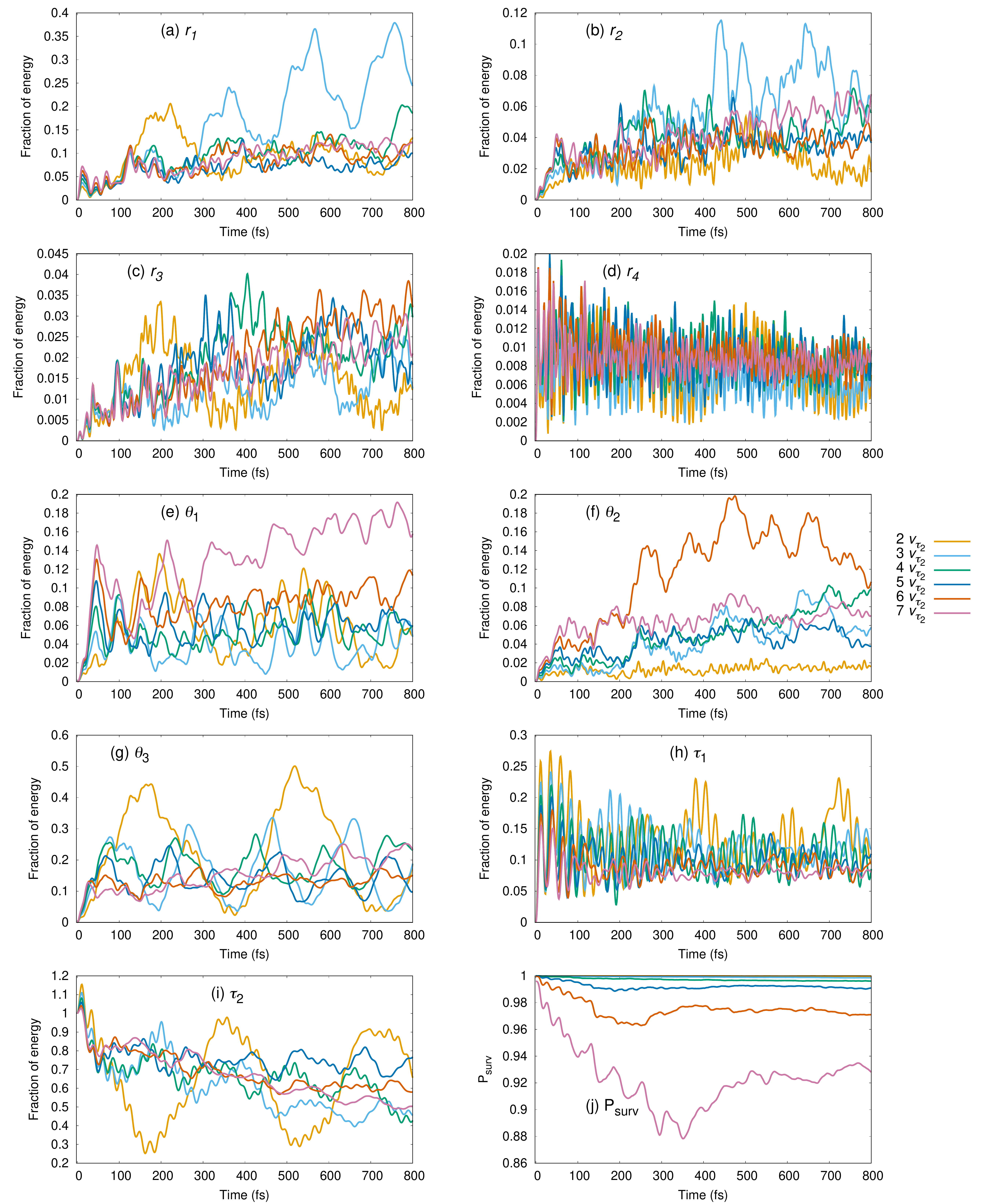} 
\caption{Fraction of energy in all nine modes ({a} to {i}) and survival probability $P_{\text{surv}}$ (j) of the propagations of overtones of local torsional mode $\tau_{2}$ of \textit{trans}-\ce{DCOOH} with 2 to 7 quanta. Note the change of scales for the y-axis for (a)-(j).}
\label{fig:energyflow_torsion_trans_DcooH}
\end{figure}

The energy flows from local torsion mode $\tau_2$ overtones in \textit{trans}-\ce{DCOOH} are depicted in Fig. \ref{fig:energyflow_torsion_trans_DcooH}. They are very similar to the \textit{trans}-HCOOH case, which confirms the minimal effect from the coupling of the $\nu_{\ce{CH 1}}$ stretch in the latter and the $\nu_{\ce{CD 1}}$ stretch here within the excitation energy range considered. Among the bright vibrational states, the coupling with the $\nu_{\ce{C-O}}$ stretching mode remains prominent especially for $3\nu_{\tau_2}$ see Fig. \ref{fig:energyflow_torsion_trans_DcooH} (a). This coupling was also observed by Nejad and Sibert.\cite{jcp_154_064301} Couplings with the dark vibrational states show similar trends with the strongest couplings arising with the $\delta_{\ce{COH 2}}$ mode again, see Fig.~\ref{fig:energyflow_torsion_trans_DcooH} (g). The involvement with the $\gamma_{\ce{CD 1}}$ rocking and $\ce{OCO-COD 2}$ deformation for excitations with 6 and 7 quanta respectively are standing out, see Figs \ref{fig:energyflow_torsion_trans_DcooH} (f) and (e), respectively. Rather than participating in the isomerisation process, these modes seem to act as energy reservoirs in the respective cases given the absence of strong periodic energy flows with these modes.

\subsection{\textit{trans}-\ce{HCOOD} $\tau_2$ torsion}

\begin{figure}
\includegraphics[width = \columnwidth]{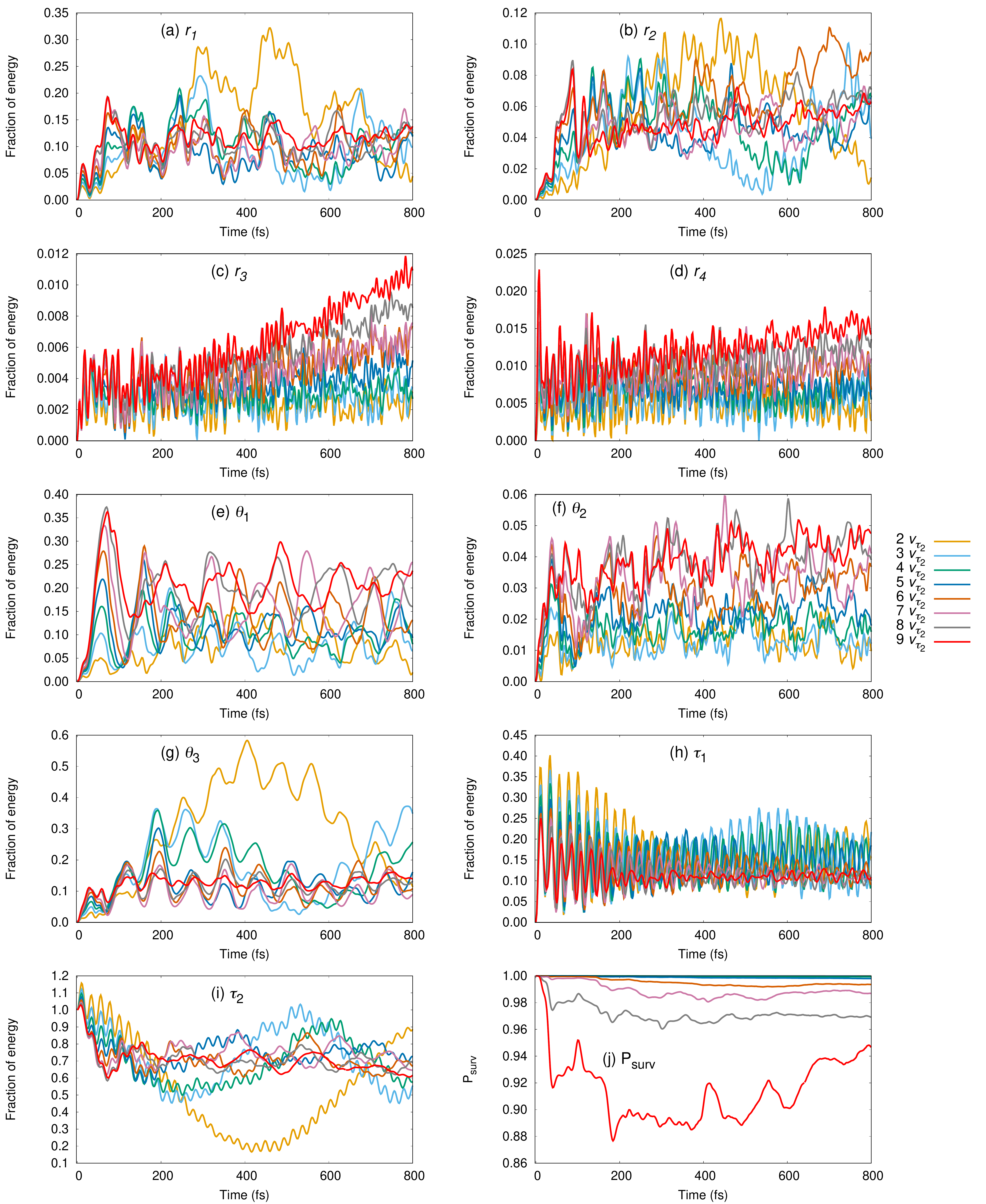} 

\caption{Fraction of energy in all nine modes ({a} to {i}) and survival probability $P_{\text{surv}}$ (j) of the propagations of overtones of local torsional mode $\tau_{2}$ of trans HCOOD with 2 to 9 quanta. Note the change of scales for the y-axis for (a)-(j).}
\label{fig:energyflow_torsion_trans_HcooD}
\end{figure}

The energy flows from local torsion mode $\tau_2$ overtones in \textit{trans}-\ce{HCOOD} are depicted in Fig. \ref{fig:energyflow_torsion_trans_HcooD}. Here, the situation relates strongly to the observations made with \textit{trans}-\ce{DCOOD}, which stresses the destructive effect of the isotopic substitution on the hydroxyl group on the coupling between the torsional mode $\tau_{\ce{COD 2}}$ and the bright $\nu_{\ce{C-O}}$ stretch. Again, no other strong coupling arises with the bright vibrational modes as in the \textit{trans}-\ce{DCOOD} case. We only point out the (most probably) accidental resonance of the first overtone with the $\delta_{\ce{COD 2}}$ dark bending mode, see Fig. \ref{fig:energyflow_torsion_trans_HcooD} (g) whereas this happened with the $\gamma_{\ce{CD 1}}$ rocking mode in trans DCOOD, see Fig. \ref{fig:energyflow_torsion_trans_DcooD} (f).

\section{Conclusions and perspectives}
We studied the IVR dynamics in formic acid and its deuterated forms up to approximatively the isomerization barrier energy. The dynamics is simulated from local mode excitations on a realistic PES using an exact kinetic energy operator in valence coordinates. The vibrational couplings are clearly identified, especially for the torsion mode leading to isomerization with particular interest for the coupled bright modes. 

An efficient indirect coupling is found between the torsion mode and the ``bright" $\ce{C-O}$ stretch mode leading to \textit{trans}-\textit{cis} isomerization, which could be exploited for laser-induced isomerization of formic acid (\ce{HCOOH}). The coupling is destroyed due to deuteration of the hydroxyl group in $\ce{DCOOD}$ and $\ce{HCOOD}$ but is preserved in $\ce{DCOOH}$. Also, there was no sign that the deuteration would cause new couplings with the torsion mode that could be used to induce an isomerization. 
This is in striking contrast with the \ce{HFCO}/\ce{DFCO} systems where new couplings with the out-of-plane mode appear when deuterated.\cite{pas07:024302}    

Motivated by the identification of the coupling of the torsion mode involved in the isomerization process with the bright $\ce{C-O}$ stretch, we plan to simulate the effect of a laser excitation on the isomerization process rather than using artificial local modes. Doing so requires the construction of the dipole moment surfaces in a body-fixed reference frame in sum of product form for use in MCTDH. We also plan to include the effect of the laser pulse on the rotational degrees of freedom explicitly, eliminating the need to consider an orientation for the molecule which has been shown to have a meaningful effect, \textit{e.g.}, in ammonia.\cite{jcp_150_014102} The kinetic energy operator will have to include explicitly the rotation and the Coriolis coupling.

\begin{acknowledgments}
The research was partly funded by the Natural Sciences and Engineering Research Council of Canada (NSERC Discovery Grant to A.B.). A.A. acknowledges the financial support of the King Baudouin Foundation, acting on behalf of the Platform for Education and Talent, FWO and F.R.S.-FNRS for his Gustave Bo\" el – Sofina Fellowship. The IISN (Institut Interuniversitaire des Sciences Nucl\'eaires) is acknowledged for its financial support.
Computational resources have been provided by the Consortium des \'Equipements de Calcul Intensif (C\'ECI), funded by the Fonds de la Recherche Scientifique de Belgique (F.R.S.-FNRS) under Grant No. 2.5020.11 and by the Walloon Region. 
We acknowledge F. Richter and P. Carbonni\`ere (Universit\'e de Pau et des Pays de l'Adour) for providing their HCOOH PES and KEOs for use in MCTDH.
\end{acknowledgments}

\underline{\textbf{AUTHOR DECLARATIONS}}\linebreak
The authors have no conflicts to disclose.

\underline{\textbf{DATA AVAILABILITY STATEMENT}}\linebreak
The data that support the findings of this study are available from the corresponding author upon reasonable request.

\bibliographystyle{apsrev4-1} 
%

\end{document}